\documentclass[fleqn,12pt]{article}
\newcommand{\AmS}{{\protect\the\textfont2
  A\kern-.1667em\lower.5ex\hbox{M}\kern-.125emS}}

\title{Neutrino Geophysics at Baksan I:
Possible Detection of Georeactor Antineutrinos}

\author{G. Domogatski$^1$, V. Kopeikin$^2$, L. Mikaelyan$^2$, V. Sinev$^2$ \\
\\
$^{1}$Institute for Nuclear Research RAS, Moscow, \\
$^{2}$Russian Research Center "Kurchatov Institute"}
 
\begin{document}

% typeset front matter
\maketitle

%\tableofcontents

\begin{abstract}
J.M. Herndon in 90-s proposed a natural nuclear fission georeactor at the center of the Earth 
with a power output of 3-10 TW as an energy source to sustain the Earth magnetic field. 
R.S. Raghavan in 2002 y. pointed out that under certain condition antineutrinos generated in 
georeactor can be detected using massive scintillation detectors. We consider the underground 
Baksan Neutrino Observatory (4800 m.w.e.) as a possible site for developments in 
Geoneutrino physics. Here the intrinsic background level of less than one event/year in a 
liquid scintillation $\sim$1000 target ton detector can be achieved and the main source of 
background is the antineutrino flux from power reactors. We find that this flux is 
$\sim$10 times lower than at KamLAND detector site and two times lower than at Gran Sasso 
laboratory and thus at Baksan the georeactor hypothesis can be conclusively tested. We also 
discuss possible search for composition of georector burning nuclear fuel by analysis of the 
antineutrino energy spectrum.
\end{abstract}

\section*{Introduction}

In this paper we consider possibilities to detect at BNO (Baksan Neutrino Observatory of Institute 
for Nuclear Research RAS) antineutrinos from georeactor using liquid scintillation spectrometer of $\sim$1000 ton target mass. The same 
spectrometer can detect $\bar{{\nu}_e}$ coming from terrestrial $^{238}$U and $^{232}$Th decays; 
the latter problem will be considered in the next publication. We mention also that here search for 
astrophysical antineutrino flux can be done.

The Earth magnetic field varies in intensity and irregularly reverses polarity with an average interval 
between reversals of about 200 000 years. This requires some variable or intermittent energy source. 
This source is understood as georeactor, i.e. as naturally varying self-sustaining nuclear chain reaction 
burning at the center of the Earth. The georeactor started $\sim$4.5 billon years ago when 
$^{235}$U/$^{238}$U enrichment was about 30\%. In the georector $^{239}$Pu is formed by 
neutron capture in $^{238}$U followed by two short-lived beta-decays: $^{238}$U$(n,{\gamma})$
$\rightarrow$ $^{239}$U(${\beta}^{-}$) $\rightarrow$ $^{239}$Np(${\beta}^{-}$) $\rightarrow$ 
$^{239}$Pu. The neutron flux in the reactor is extremely low and, in contrast with man-made high 
flux Power reactors, $^{239}$Pu does not contribute to the fission power and decays in $^{235}$U: 
$^{239}$Pu(${\alpha}, T_{1/2} = 2.4\times 10^{4}$ y) $\rightarrow$ $^{235}$U. Thus the 
georeactor operates in a breeder regime and reproduces $^{235}$U through $^{238}$U
$\rightarrow$ $^{239}$Pu $\rightarrow$ $^{235}$U cycle. An average thermal power output of the 
Uranium based reactor is assumed to amount of 3$-$6 TW. Had Thorium been included the power 
could be higher. Variations of georeactor power originate from self-poisoning due to accumulation 
of fission products and subsequent removal of these products by diffusion or some other mechanism. 
This is a short and very schematic summary of georeactor concept proposed in a number of 
publications by J.M. Herndon. [1].

Nuclear fission chain reaction can occur in nature. In 1956 P. Kuroda showed that thick seams of 
uranium ore might, 2 billion years ago, have been able to support chain reactions and function 
as a natural nuclear reactors [2]. 16 years later remains of a natural nuclear fission reactor were actually 
found in the mine at Oklo in the Republic of Gabon in Africa [3]. 

Herndon's idea about georeactor located at the center of the Earth, if validated, will open a new era 
in planetary physics. However it is not clear whether further geophysical, chemical etc studies can in 
foreseeable future give a decisive confirmation (or disproof) of this reactor. Particle physics can give 
another approach to the problem. In 2002 y R.S. Raghavan pointed out that under certain conditions 
a direct and conclusive test could be obtained by detection of antineutrinos from georeactor [4]. 

Below we consider: 
\begin{itemize}
\item Georeactor: expected $\bar{{\nu}_e}$ rate and spectrum.
\item Detector design and backgrounds.
\end{itemize}

In the last section we compare $\bar{{\nu}_e}$ energy spectra emitted in $^{235}$U, $^{238}$U
and $^{233}$U fission and discuss possibilities to search for georeactor fuel composition using 
$\bar{{\nu}_e}$ spectroscopy.

\section{Georeactor: expected $\bar{{\nu}_e}$ rate and spectrum}

Georeactor antineutrinos are detected in liquid scintillation spectrometer via the inverse beta-decay 
reaction 

\begin{equation}
\bar{{\nu}_e}+p \rightarrow n + e^{+}
\end{equation}
 
The visible positron energy $E_e$ is related to the $\bar{{\nu}_e}$ energy as

\begin{equation}
E_{e} = E - 1.80 + E_{annihil} - r_{n} \approx E - 0.8,
\end{equation}

where 1.80 MeV is the threshold of the reaction and $r_n$ is the neutron recoil energy.
The signature of neutrino event is $e^{+}$ and 2.2 MeV neutron signals correlated in time and space. 

Calculated antineutrino interaction rate $N_{{\nu}GR}$ for georeactor power $W$ =3$-$10 TW and 
$N_{p} = 10^{32}$ target protons $N_{{\nu}GR}$ = (33$-$110) / year is found for no-oscillation 
case and detection efficiency $\epsilon$ = 100\%, the Earth radius $R_{Earth}$ = 6370 km and 
typical PWR reactor parameters: 
\begin{eqnarray}
N_{{\nu}GR} \approx (33-110) /{\rm year \ with \ } 10^{32} {\rm protons, \ 3-10 \ TW}, \\ \nonumber
{\epsilon} = 100\%  \ {\rm and \ no \ oscillation},
\end{eqnarray}

which is exactly what has been found in [4]. Had $^{235}$U neutrino fission parameters been used, the 
rate would be $\sim$10\% higher.

Positron visible energy spectrum is shown in Fig. 1. 

\section{Detector design and backgrounds}

The sensitivity of low energy antineutrino detection depends on detector size and level of background. 
In the past 10 years the sensitivity was increased, in two steps (CHOOZ, KamLAND), by a factor of 
$\sim10^8$ and approached $\sim$1 event per year per $\sim$1000 ton LS target.

The main features of future BNO detector design and location can be:

a) Three-concentric zone detector design (Fig. 2). The central $\sim$14 m diameter zone one is $10^{32}$
 H atom liquid scintillator target contained in a spherical transparent balloon. Zone two is a buffer of 
non-scintillation oil contained in a $\sim$19 m diameter stainless steel vessel; on the inner surface of the 
vessel are mounted PMTs with $\sim$30\% photo-cathode coverage. A transparent acrylic barrier 
protects radon emanations from penetrating in the LS of zone one. The zone three is $\sim$22 
m-diameter water Cherenkov detector which protects the inner parts from neutrons and 
$\gamma$-rays coming from the surrounding rock and gives veto signals for cosmic muons.

b) Deep underground position of the detector to reduce muon-induced backgrounds. BNO is located 
at the site with 4 800 mwe rock overburden, which is much deeper than KamLAND's 2700 mwe 
position.

c) Highest purification of zone 1 (LS) and zone 2 (oil) (U, Th and K concentrations as low as $10^{-17}$ 
g/g).

Experience accumulated in KamLAND experiment [5] shows that with a) - c) conditions intrinsic 
detector background at BNO of less than 1/year in a LS target with $10^{32}$ H atoms can be achieved. 

Most important condition for successful detection of georeactor antineutrinos is not too high 
antineutrino flux coming from Power reactors. Using data from [6] the $\bar{{\nu}_e}$ interaction 
no-oscillation rate $N_{{\nu}PWR}$ is (see Table):

\begin{table}[htb]
\caption{Antineutrino backgrounds at BAKSAN from Power reactors}
\label{table}
\vspace{10pt}
\begin{tabular}{c|c|c|c|c|c}
\hline
Country & Number & Thermal & Distance & Energy & Rate$^{*}$ \\
or Plant & of cores & Power, GW & from BNO, & flux$^{*}$, & per 10$^{32}$ p \\
        & & & km &  J/cm$^2$/year & year$^{-1}$ \\
\hline
Rostov & 1 & 3 & 463 & 2.99 & 5.34  \\
Kursk	& 4& 12.8	& 1070	& 2.38	& 4.26	\\
Smolemsk	&	3	&	9.6	&	1500	&	0.91	&	1.6	\\
Balakovo	&	4	&	12	&	1035	&	2.37	&	4.27	\\
Tver	&	2	&	6	&	1600	&	0.5	&	0.89	\\
Novovoronezh	&	3	&	5.75	&	945	&	1.37	&	2.46	\\
Rovno	&	3	&	5.75	&	1550	&	0.51	&	0.9	\\
Khmelnitsky	&	1	&	3	&	1395	&	0.33	&	0.59	\\
Chernobyl	&	1	&	3.2	&	1278	&	0.39	&	0.7	\\
Zaporozhie	&	6	&	18	&	612	&	10.3	&	18.33	\\
Yuzhno- 	&	3	&	9	&	1035	&	1.79	&	3.2	\\
ukrainskaya & & & & & \\
Great 	&	35	&	38.5	&	&	0.71	&	1.28	\\
Britain & & & & & \\
France	&	58	&	204.8	&	&	5.05	&	9.04	\\
Germany	&	19	&	69.5	&		&	2.28	&	4.07	\\
Baltic &	&	69.7	&&	2.68	&	4.79	\\
countries & & & & & \\
Nearest & & 62.4 &	&	2.63	&	4.7	\\
European & & & & & \\
countries & & & & & \\
Armenia	 &	1	&	1.375	&	400	&	1.83	&	3.28	\\
Bucher$^{**}$ & 1 & 3	&	1760	&	0.21	&	0.37	\\
Pakistan	&	1	&	0.375	&	3130	&	0.01	&	0.017	\\
India	&	10	&	5.8	&	4320	&	0.08	&	0.14	\\
\hline
Total	&		&	&		& 39.35	 & 70.5	\\
\hline
\end{tabular}\\[2pt]
{\small $^{*}$ Average power is assumed 0.85 of its maximal value.}\\
{\small $^{**}$ Bucher Power Plant is under construction now.}
\end{table}

\begin{eqnarray}
N_{{\nu}PWR} = 70.5 /{\rm year \ with \ } 10^{32} {\rm protons}, \\ \nonumber
{\epsilon} = 100\%  \ {\rm and \ no \ oscillation},
\end{eqnarray}

This rate is $\sim$10 times smaller than at Kamioka site and two times smaller than at Gran Sasso 
(For KamLAND and Gran Sasso data see ref. [4]). Using known PWR powers and their distances 
from BNO this rate can be calculated with $\sim$3\% systematic uncertainty. 

Antineutrino interaction rates (3, 4) are obtained for no oscillation case and 100\% detection efficiency. 
With realistic ${\epsilon}$ = 80\% and LMA oscillation parameters the detection rates are two times 
lower. Nevertheless in $\sim$2 years of data taking a 3 TW georeactor can be conclusively confirmed.

\section{On analysis of fuel composition in georeactor}

Imagine that the georeactor hypothesis is confirmed. The next step could be efforts to obtain direct 
information on composition of the nuclear fuel, which no doubt, would be of primary geophysical 
importance. 

The shape of reactor $\bar{{\nu}_e}$ energy spectrum depends on contributions of fissile isotopes to 
the total chain reaction rate. Thus measurement of the $\bar{{\nu}_e}$ spectrum provides information 
on nuclear fuel composition. This idea was first proposed years ago [7] and later was confirmed in 
experiments at reactors [8].

In water-cooled thermal neutron power reactors with (initial) $^{235}$U/$^{238}$U enrichment 
$\sim$4\% fast neutron fission of $^{238}$U contributes typically 7.5\% to the total reactor fission 
rate. In the fast neutron georeactor the $^{238}$U contribution can be expected to be much higher 
(no information on this subject is given in [1]). Calculated ratio of reaction (1) positron spectra 
induced by $^{238}$U and $^{235}$U fission antineutrinos (Fig. 3) considerably departs from 
unity. Thus, using shape analysis and with larger statistics, contribution of $^{238}$U fission can be estimated. 

We continue speculations on the georeactor nuclear fuel composition. Suppose that initially ($\sim$4.5 
billion years ago) large amount of $^{232}$Th was present in the georeactor core. Then $^{233}$U is 
formed through neutron capture and two beta decays: 
$^{232}$Th$(n, {\gamma}) \rightarrow ^{233}$Th(${\beta})  \rightarrow ^{233}$Pa(${\beta}) \rightarrow ^{233}$U.
$^{233}$U with its large fission cross section would largely contribute to the total georeactor fission 
rate. 

We have calculated the $^{233}$U fission $\bar{{\nu}_e}$ energy spectrum (V. Kopeikin et al., to 
be published) and found that it is much softer than $^{235}$U fission $\bar{{\nu}_e}$ energy 
spectrum (Fig. 3). Thus, if contribution of $^{233}$U fission is sufficiently large, this can be found 
in experiments considered here. We note also that if $^{233}$U and $^{238}$U equally contribute 
to georeactor fission power, the resulting positron spectrum can look very much like that of $^{235}$U.

\section*{Conclusions}

Hypothesis of 3 TW georeactor burning inside the Earth can be conclusively tested at Baksan 
with a few years of data taking using $\sim$1000 target ton liquid scintillation detector. With longer 
time/larger LS mass a search for dominant nuclear fuel components can be done.

\section*{Acknowledgments }
We are grateful to Professors O.G. Ryazhskaya, J.M. Herndon and Yu. Kamyshkov for fruitful 
discussions. This study is supported by RFBR grant 03-02-16055 and Russian Federation 
President's grant 1246.2003.2.

\appendix

\end{document}